\documentclass[journal]{IEEEtran}
\ifCLASSINFOpdf
\else
\fi

\usepackage{array}
\usepackage{comment}
\usepackage{amsmath}
\usepackage{amssymb}
\usepackage{setspace}
\usepackage{graphicx}
\usepackage[tight,footnotesize]{subfigure}
\usepackage{makecell,multirow,diagbox}
\usepackage[justification=centering]{caption}
\usepackage{epstopdf}
\usepackage{algorithm}
\usepackage{algorithmicx}
\usepackage{algpseudocode}
\usepackage{subfigure}
\usepackage{float}
\usepackage{cite}
\begin{document}

\title{A Large-scale Concurrent Data Anonymous Batch Verification Scheme for Mobile Healthcare Crowd Sensing}

\author{Jingwei~Liu,~\IEEEmembership{Member,~IEEE,} Huijuan~Cao, Qingqing~Li, Fanghui~Cai,
        \\Xiaojiang~Du,~\IEEEmembership{Senior Member,~IEEE} and Mohsen Guizani,~\IEEEmembership{Fellow,~IEEE}
\thanks{Jingwei Liu, Huijuan Cao, Qingqing Li, and Fanghui Cai are with State Key Laboratory of Integrated Services Networks, Xidian University, Xi'an, 710071, China. (e-mail: jwliu@mail.xidian.edu.cn, caohuijuan345@163.com, 15229259171@163.com, caidoreen@163.com.)}
\thanks{Xiaojiang Du is with the Department of Computer and Information Sciences, Temple University, Philadelphia, PA 19122, USA. (e-mail: dxj@ieee.org.)}
\thanks{Mohsen Guizani is with the Department of Electrical and Computer Engineering, University of Idaho, Moscow, Idaho, USA. (e-mail: mguizani@ieee.org.)}
}

\maketitle

\begin{abstract}
Recently, with the rapid development of big data, Internet of Things (IoT) brings more and more intelligent and convenient services to people's daily lives. Mobile healthcare crowd sensing (MHCS), as a typical application of IoT, is becoming an effective approach to provide various medical and healthcare services to individual or organizations. However, MHCS still have to face to different security challenges in practice. For example, how to quickly and effectively authenticate masses of bio-information uploaded by IoT terminals without revealing the owners' sensitive information. Therefore, we propose a large-scale concurrent data anonymous batch verification scheme for MHCS based on an improved certificateless aggregate signature. The proposed scheme can authenticate all sensing bio-information at once in a privacy preserving way. The individual data generated by different users can be verified in batch, while the actual identity of participants is hidden. Moreover, assuming the intractability of CDHP, our scheme is proved to be secure. Finally, the performance evaluation shows that the proposed scheme is suitable for MHCS, due to its high efficiency.
\end{abstract}


\IEEEpeerreviewmaketitle

\section{Introduction}

\IEEEPARstart{I}{ot}, as a promising paradigm, can change the interactive way between networks and the physical world \cite{rajkumar2010cyber}. Meanwhile, with the popularization and development of wireless sensors, a new perceptual architecture - mobile crowd sensing (MCS) \cite{ganti2011mobile, guo2014participatory}, has emerged. It provides a important technical support for the integration of the physical world with higher layer applications in IoT. As an important application branch of MCS, mobile healthcare crowd sensing (MHCS) provides more convenient medical and healthcare services for organizations or individual.

\begin{figure}[ht]
  \centering
  \includegraphics[width=9cm]{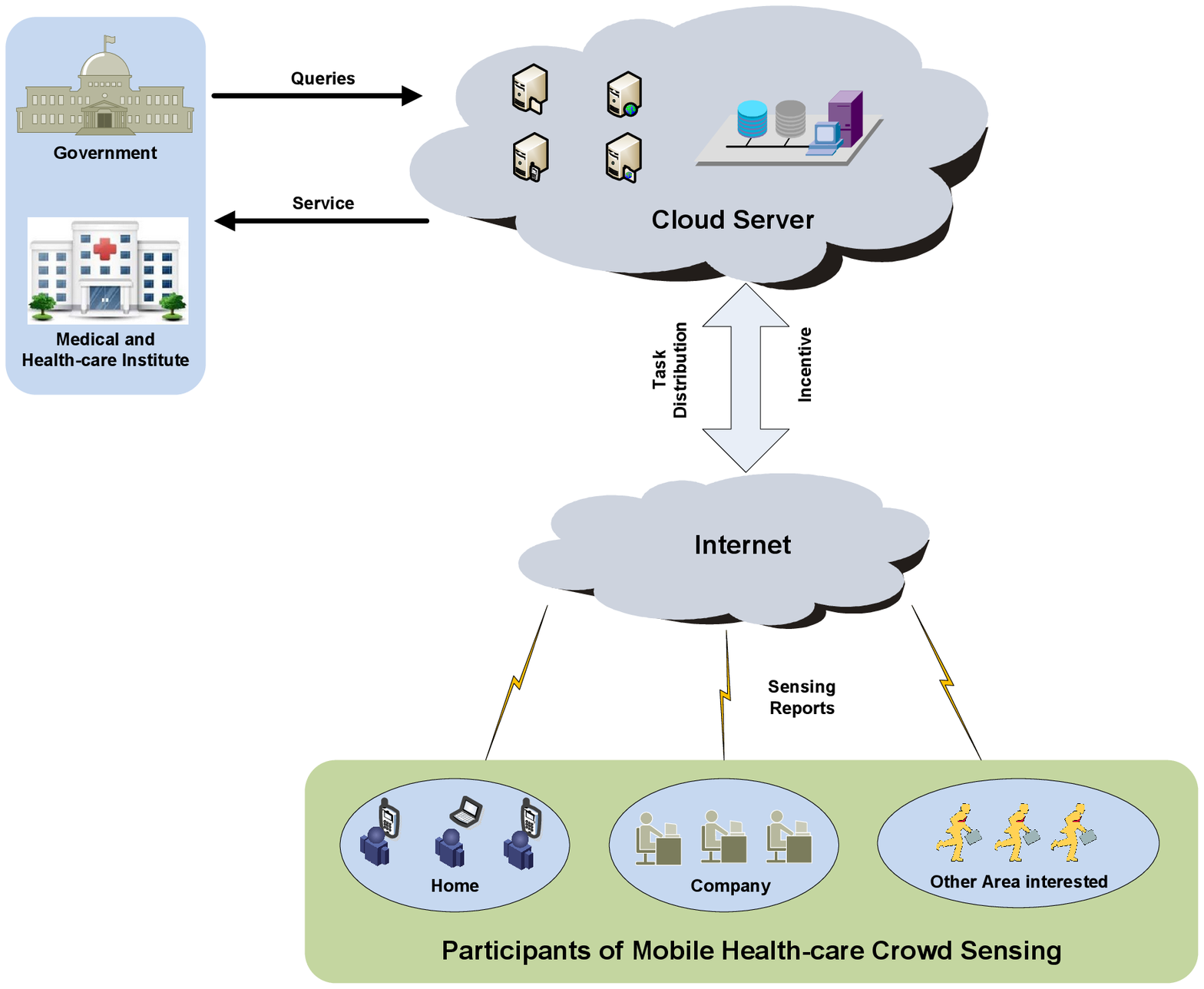}\\
  \caption{A simple architecture of the MHCS system}
  \label{MHCS_archi}
\end{figure}

Mobile healthcare crowd sensing (MHCS), combining the merits of mobile crowd sensing with remote healthcare, is becoming a research hotspot. On one hand, participants in MCS upload health data collected by mobile terminals to cloud server and enjoy various services by healthcare organizations. On the other hand, remote healthcare system can provide health information and medical service anytime and anywhere, by analyzing the individual health data and patient vital signs submitted to remote health apps installed in mobile terminals or monitoring devices. Therefore, MHCS can not only provide real-time medical services to individual or community, but also improve the ability of healthcare organizations to monitor, track and control certain diseases on some regions.

However, there are still many security threats and privacy issues in MHCS: a) the collected health data may deduce users' sensitive information, such as identity, personal activities and health status; b) the data may be obtained or changed by an opponent, which will bring damage to people's health and property, even people's lives; c) these data collected by mobile devices should be processed safely in a real-time manner, otherwise the quality of medical service will be reduced. Therefore, the security and privacy preservation for MHCS is need to be considered emergently. So, more and more privacy-preserving schemes \cite{wang2013artsense, lane2010survey, boneh2003aggregate, zhu2016efficient, shao2016threshold, liu2016efficient, bao2015new, wang2016rescuedp, zhuo2016privacy, wang2014oruta, horng2015efficient, Shim2012, liu2014certificateless, zhang2016privacy, zhu2016secure, yuan2016privacy, wang2016catch} have been proposed in recent years. In this work, we also mainly focus on the privacy preservation for MHCS.

According to \cite{lane2010survey}, a simple architecture of the MHCS system consists of MHCS participants, a cloud sever, and healthcare organizations. In MHCS, as shown in Fig. \ref{MHCS_archi}, the cloud sever publishes sensing task for specific purpose. The participants receive a sensing task published from cloud sever, then they collect and upload the relevant health data to the sever. Meanwhile, the cloud sever will deliver the requested information to specific organizations or healthcare institutes so as to make further analysis. However, millions of participants submit numerous biomedical data to the cloud sever, which will lead to data transmission obstacles and storage capacity burdens. An efficient approach named aggregate signature (AS) can improve the efficiency of the verification on numerous signatures and reduce the overhead of storage and bandwidth. The first AS scheme based on traditional public key cryptography was proposed by Boneh et al. \cite{boneh2003aggregate} in 2003. It allowed multiple users to generate the signatures on different messages respectively and verify them in batch.

Following Boneh's work, many AS schemes were proposed subsequently, but most of them were involved in the complicated certificate management problem. Thus, certificateless public key cryptosystem (CL-PKC) appeared to solve this issue. In 2007, Castro and Dahab \cite{castro2007efficient} first introduced the concept of certificateless aggregate signature (CL-AS) that combined the merits of aggregate signature with CL-PKC. Then, Gong et al. raised the formal security model for CL-AS in \cite{gong2007two} in the same year. After the initial work, lots of CL-AS have been proposed \cite{tu2014reattack, malhi2015efficient, zhang2009new, xiong2011strong}.

In this paper, we put forward a large-scale concurrent data anonymous batch verification scheme for MHCS. The main work of this paper are summarized as follows:

\begin{itemize}
\item
The proposed scheme can provide bio-information batch verification and anonymous authentication for MHCS systems.

\item
Based on the hardness of the Computational Diffie-Hellman Problem (CDHP), it is formally proved that our scheme is secure against the existential forgery attack on adaptively chosen message.

\item
In the quantitative performance evaluation, our scheme achieves less computation overhead compared with the previous schemes. It is very suitable for the MHCS systems in practice.
\end{itemize}

The rest part of this paper is organized as follows. Firstly, we introduce the reference model, security model and design goals in Section II. In Section III, we improve a CL-AS scheme with the security proof. In Section IV, we describe the the large-scale concurrent data anonymous batch verification scheme in detail. In Section V, we analyze the performance. Finally, we conclude this paper in Section VI.

\section{MODELS AND DESIGN GOALS}
For a better understanding, we first put forward the relevant models for MHCS, and then raise design goals.

\subsection{Reference Model}
\begin{figure}
  \centering
  \includegraphics[width=9cm]{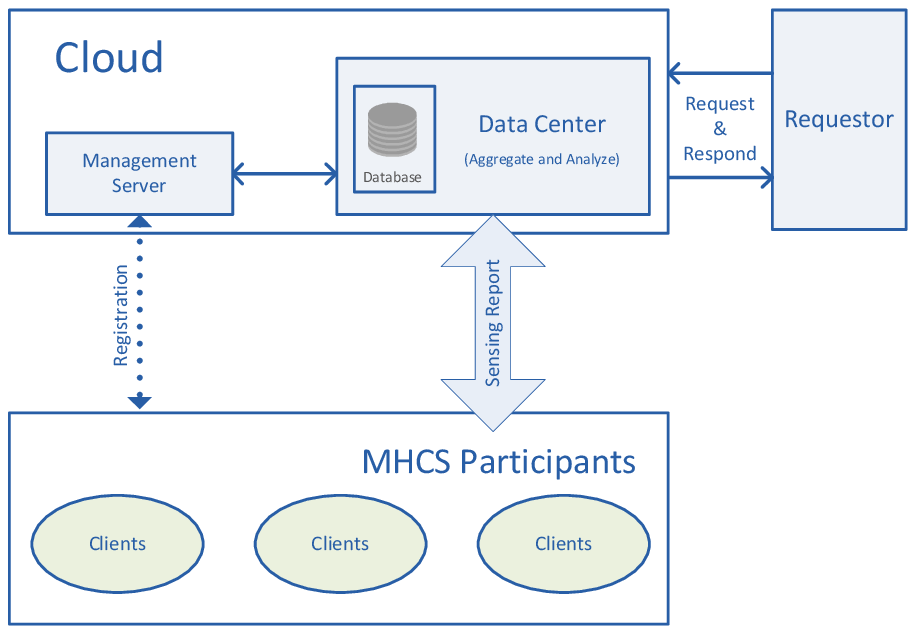}\\
  \caption{The reference model for MHCS systems}\label{reference_model}
\end{figure}

The reference model for MHCS scenarios consists of four entities: Requestor, Data Center (DC), Management Server (MS), and MHCS Participants, as shown in Fig.\ref{reference_model}.
\begin{itemize}
\item \textbf{Requestor:} The requestors can submit healthcare sensing tasks to DC for some specific purposes. And they can further analyze the final report from DC to predict certain medical or health issues in some regions.
\item \textbf{Data Center (DC):} It can publish and manage healthcare sensing tasks according to the demands of the requestors. Also, it is responsible for aggregating and verifying all collected health data from different participants.
\item \textbf{Management Server (MS):} MS is a trusted third party who can manage the participants' registration information in MHCS systems. It is in charge of issuing the a half private keys for legitimate participants and distributes the index of the participants to cover their actual identity. Here, DC can use the index to authenticate the uploaded health data from the participants.
\item \textbf{MHCS Participants:} MHCS Participants refer to the mobile clients who collect and submit relevant health data using smart terminals for Data Center (DC).
\end{itemize}

\begin{table*}[ht]
\begin{center}
\footnotesize
\caption{Notations}\label{tabl:notation}
\setlength{\extrarowheight}{0.25cm}
\begin{tabular}{c|c||c|clllll}
\hline
 $G_a$ & a cyclic additive group of order $q$ & $P$ & a generator of $G_a$ \\
 $G_m$ & a cyclic multiplicative group of order $q$ & $e(\cdot)$ & a bilinear map: $G_a \times G_a \rightarrow G_m$ \\
 $\sigma_i$ & digital signature of the participant with $ID_i$ & $m_i$ & healthcare data of the participant with $ID_i$\\
 $\sigma$ & aggregate signature &$A_i$ & An adversary on type $i$\\
 $Q_{DC}$  &  DC's public key &$s_{DC}$ & DC's private key\\
 $Q_{MS}$  &  MS's public key  &  $s_{MS}$ & MS's private key \\
 $\langle Q_{1i},Q_{2i}\rangle$ & the public key of the participant with $ID_i$  &  $\langle s_{1i},S_{2i}\rangle$  &  the private key of the participant with $ID_i$  \\
 $l$  &  system security parameter & $q$ & a large prime number \\
 $\textsf{$H_1$}(\cdot)$ & a hash function: ${\{0, 1\}}^\ast\times G_a\rightarrow G_a$ &
 $\textsf{$H_2$}(\cdot)$ & a hash function: ${\{0, 1\}}^\ast\times G_a \rightarrow Z^*_q$\\
 \hline
\end{tabular}
\end{center}
\end{table*}

\subsection{Security Model}

As security issues studied in \cite{wu2014security,wu2014mobifish,huang2014achieving,du2008security}, design of multi-party mobile computing scheme requires extra caution on security and privacy issue. To make better security analysis, we refer to the security model defined in \cite{huang2011certificateless}, in which there are two types of opponents who are able (or unable) to replace certain participants' public key without (or with) the management server's private key. In this model, it can be proved that our scheme is secure against the above two kinds of opponents, if the following computational Diffie-Hellman problem (CDHP) is intractable. Here, we give the definition of the CDHP: in a large prime order $q$ cycle additive group $G_a$, $\forall xP, yP$ with a generator $P$ and unknown $x,y\in Z^*_q$, get $xyP$ finally.

\subsection{Design Goals}

Our design goals aim at designing a large-scale concurrent data anonymous batch verification scheme for MHCS, which achieves following properties:

\begin{itemize}
\item \textbf{Batch authentication:} The authentication information in the signed bio-data from large-scale MHCS participants could be aggregated and verified effectively by DC.
\item \textbf{Non-repudiation:} MHCS participants cannot deny that they have submitted the related health data to DC.
\item \textbf{Anonymity:} Although DC can acquire and check the aggregated authentication message, it cannot obtain the real identity of the data provider.
\end{itemize}

\section{An Improved CL-AS Scheme}
Key management is essential for security \cite{du2009transactions,xiao2007survey,du2007effective}.To provide a cryptographic essential for our design goals, we primarily propose an improved CL-AS scheme and then give the relevant security proof in this section. It can not only be used to realize batch verification, but also can deal with the key escrow problem of identity-based public key cryptosystem (ID-PKC)\cite{shamir1984identity}. Due to these merits, it could be the key to designing a large-scale concurrent data anonymous batch verification scheme for mobile healthcare crowd sensing systems. Before describing the new certificateless aggregate signature scheme, we first introduce the concept of bilinear pairing.

\subsection{Bilinear Pairing}

A bilinear pairing map, formally defined as $e: G_a\times G_a\to G_m$, should satisfy the following three properties, in which $G_a$ is a additive group, $G_m$ is a multiplicative group, $q$ is the order, $P$ is the generator of $G_a$.

\begin{itemize}
\item Bilinear: $\forall L, M, N \in G_a$, $e(L, M+N) = e(L, M)e(L,$ $N)$ or $\forall x, y\in Z^*_q$, $e(xM, yN) = e(M, N)^{xy} = e(xyM,$ $N) = e(M, xyN)$;
\item Non-degenerate: $\exists M, N \in G_a$, satisfy $e(M, N)\not=g$. Here, $g$ is the generator of $G_m$;
\item Computable: $e(\cdot, \cdot)$ should be efficient, $\forall M, N \in G_a$.
\end{itemize}

\subsection{Design of the new CL-AS Scheme}

In this part, the detailed CL-AS scheme is constructed. We give the specification on \emph{Setup}, \emph{Set-Partial-Key}, \emph{Signing}, \emph{Verification}, \emph{Aggregation}, and \emph{Aggregate Verification}, described as follows:

\begin{itemize}
\item[1)] \emph{Setup:} Key Generation Center (KGC) initializes and establishes the system as follows:
 \begin{itemize}
  \item[a.] Construct two cyclic groups $(G_a, +)$ and $(G_m, \cdot)$ with additive operation and multiplicative operation respectively. Their order is a secure large prime $q$ meeting a security parameter $l$. Set a pairing operator, $e: G_a \times G_a \rightarrow G_m $ that satisfies the properties described above. Then, select two secure hash functions $H_1 : {\{0, 1\}}^\ast\times G_a \rightarrow G_a$ and $H_2 : {\{0, 1\}}^\ast\times G_a \rightarrow Z^*_q$.
   \item[b.] Key Generation Center (KGC) picks a random number $s_{KGC}\in Z^*_q$ for $Q_{KGC} = s_{KGC}P$. Here, $\langle s_{KGC}, Q_{KGC} \rangle$ is its private/public key pair. Then, KGC publishes $\langle l, q, P, G_a, G_m, e, H_1, H_2, Q_{KGC} \rangle$ as the system parameters, while store $s_{KGC}$ as its private key secretly.
 \end{itemize}
\item[2)] \emph{Set-Partial-Key:} It consists of two part algorithms, one is to generate the partial key by a client or a signer, the other one is to compute the partial key by the KGC.
  \begin{itemize}
   \item[a.] A client or a signer, marked as $C_i$, obtains his or her partial secret key by choosing $s_{1i}\in Z^*_q$ randomly and the partial public key by computing $Q_{1i} = s_{1i}P$.
   \item[b.] $C_i$ sends his/her $id_i$ to KGC and request the partial key for the identity $id_i$. KGC calculates $Q_{2i} = H_1(id_i,Q_{1i})$, $S_{2i}=s_{KGC}Q_{2i}$ for it and distributes the half private key to $C_i$ through secure channels. Hence, $C_i$ can obtain the public key $\langle Q_{1i},Q_{2i} \rangle$ and the private key $\langle s_{1i}, S_{2i} \rangle$. Note that, each identity only can be used once.
   \end{itemize}
\item[3)] \emph{Signing:} The signer chooses $k_i\in Z^*_q$ randomly and then sign a message $m_i$, as follows:
\begin{equation} \label{eqn1}
  \begin{split}
  V_i &= k_iQ_{1i} \\
  h_i &= H_2(m_i,V_i) \\
  U_i &= S_{2i}+k_ih_is_{1i}Q_{KGC}
  \end{split}
  \end{equation}

Then, the signer view the pair $\sigma_i = \langle V_i,U_i \rangle$ as the signature on $m_i$.

\item[4)] \emph{Verification:} To ensure the validity of the signature $\sigma_i$ signed by a $C_i$ on the message $m_i$, the verification procedure is as follows:
\begin{equation} \label{eqn2}
  \begin{split}
  Q_{2i}  &=H_1(id_i, Q_{1i}) \\
  h_i     &=H_2(m_i, V_i) \\
  e(U_i,P)&=e(Q_{2i}+h_iV_i,Q_{KGC})
  \end{split}
  \end{equation}

Obviously, if the above equations hold, the signature is valid. Additionally, the proposed scheme also satisfies correctness:
\begin{equation*}
    \begin{array}{rcl}
        e(U_i,P)&= &e(S_{2i}+k_ih_is_{1i}Q_{KGC},P)\\
                &= &e(S_{2i},P)e(k_ih_is_{1i}Q_{KGC},P)\\
                &= &e(s_{KGC}Q_{2i},P)e(k_ih_is_{1i}P,Q_{KGC})\\
                &= &e(Q_{2i},s_{KGC}P)e(k_ih_iQ_{1i},Q_{KGC})\\
                &= &e(Q_{2i},Q_{KGC})e(h_iV_i,Q_{KGC})\\
                &= &e(Q_{2i}+h_iV_i,Q_{KGC})
    \end{array}
\end{equation*}

\item[5)] \emph{Aggregation:} To obtain the final signature from all $\sigma_i$ of the message $m_i$, the aggregator computes in the following way:
\begin{equation} \label{eqn3}
  \begin{split}
  U   &= \sum_{i=1}^n U_i \\
  V   &= \sum_{i=1}^n h_iV_i \\
  Q_2 &= \sum_{i=1}^n Q_{2i}
  \end{split}
  \end{equation}

The $\sigma =\langle U,V \rangle$ is the final aggregated signature.

\item[6)] {Aggregate Verification:} On receiving an aggregate signature $\sigma$ for aggregating $C_i$ (from $i=1,2,\ldots, n$) and the public key $Q_{KGC}$, the verifier will authenticate the aggregate signature. And the signature $\sigma$ can be authenticated correctly, if the integrated formula holds: $e(U,P)= e(Q_2+V,Q_{KGC})$. Here, we give the proof of the equation on its correctness as follows:
  \begin{equation*}
    \begin{array}{rcl}
        e(U,P) & = & e(\sum\limits_{i=1}^{n} U_i,P)\\
                   & = & \prod\limits_{i=1}^{n}e(U_i,P)\\
                   & = & \prod\limits_{i=1}^{n}e(S_{2i}+k_ih_is_{1i}Q_{KGC},P)\\
                   & = & \prod\limits_{i=1}^{n}e(S_{2i},P)e(k_ih_is_{1i}Q_{KGC},P)\\
                   & = & e(Q_2,Q_{KGC})e(V,Q_{KGC})\\
                   & = & e(Q_2+V,Q_{KGC})
    \end{array}
  \end{equation*}
\end{itemize}

\subsection{Security Proof}
To make it convincing, it is proved that the proposed CL-AS scheme is existentially unforgeable against adaptively chosen message attacks in the random oracle model if the CDHP is intractable. As described in section II, two types adversary, named $\mathcal{A}_1$ and $\mathcal{A}_2$ who attempt to forge a legal signature with different abilities (able/unable to use the PKC's private key). We will prove the security of the proposed CL-AS under $\mathcal{A}_1$ and $\mathcal{A}_2$'s attacks respectively. The detailed proofs are as follows:

\textbf{Theorem 1.} If the adversary $\mathcal{A}_1$ could break the proposed scheme by making $q_{1/2}$ queries to $H_{1/2}$, $q_k$ queries to Extract-Queries, $q_s$ queries to Secret-Key-Queries, $q_p$ queries to Public-Key-Queries, $q_r$ queries to Replace-Public-Key queries, and $q_{sig}$ to CLAS-Sign-Queries, so CDHP could be solved within:
            $$t+(q_1+q_2+q_k+q_s+q_p+q_r+q_{sig})t_m$$
with probability:
$$\varepsilon'  \geq \frac{1}{(q_k+1)e}\varepsilon$$

\textbf{Proof.} Let $\mathcal{C}$ be a challenger trying to solve a CDHP instance $(P, aP, bP)$ in $G_a$. For $a, b\in Z_q^*$, we set $X = aP$ and $Y = bP$. $X, Y\in G_a$ is a CDHP instance in $G_a$. $\mathcal{A}_1$ interacts with $\mathcal{C}$ as the model in \cite{malhi2015efficient}. $\mathcal{C}$ sets $Q_{KGC} = X$. Suppose $\mathcal{A}_1$ is a PPT Turing machine taking only open data as input,who has a advantage to break the proposed CLAS scheme with non-negligible probability. Given two random oracles which are $H_1$ and $H_2$ respectively, $\mathcal{C}$ gives the parameters $\langle l, q, P, G_a, G_m, e, H_1, H_2, Q_{KGC} \rangle$ to $\mathcal{A}_1$. $\mathcal{C}$ tries to simulate all above oracles to obtain the valid signatures of any message $m_i$ as the real signer. List $L = \langle id_i, s_{1i}, S_{2i}, Q_{1i} \rangle$ is maintained by $\mathcal{C}$. Throughout the proof process, $\perp$ means the value of a variable is invalid. In particularly, $\mathcal{A}_1$ can query as follows:

\begin{itemize}
\item
\textbf{H$_1$-Queries:} On receiving a query $I_i = \langle id_i, Q_{1i} \rangle$ on $H_1$ from $\mathcal{A}_1$, with a list of tuple $\langle I_i, c_i,\alpha_i, Q_{2i} \rangle$, called $L_c$, $\mathcal{C}$ can simulate oracle $H_1$ as follows:
\begin{itemize}
\item[-]
If $I_i$ already exists in $L_c$, $\mathcal{C}$ outputs related $Q_{2i}$.
\item[-]
Otherwise, $\mathcal{C}$ sets $c_i=0$ with probability $\lambda$ and $c_i = 1$ with probability $(1-\lambda)$. If $c_i = 0$, $\mathcal{C}$ chooses $\alpha_i \in Z_q^*$ and outputs $Q_{2i} = b\alpha_iP\in G_a$. If $c_i = 1$, then $Q_{2i} = \alpha_iP$. In both cases, $\mathcal{C}$ inserts a tuple $\langle I_i, c_i,\alpha_i, Q_{2i} \rangle$ to $L_c$.
\end{itemize}
\item
\textbf{H$_2$-Queries:} $\mathcal{C}$ simulates $H_2$ by maintaining a list $L_v$ with $\langle J_i, h_{i}\rangle$. Here, $J_i = \langle m_i,V_i \rangle$, $m_i \in {\{0, 1\}}^\ast$ and $V_i \in G_a$. On inputting $J_i$ to $H_2$, $\mathcal{C}$ does as follows:
\begin{itemize}
\item[-]
If $J_i$ already exists in $L_v$, $\mathcal{C}$ outputs the same answer.
\item[-]
Otherwise, $\mathcal{C}$ chooses $h_{i}\in Z_q^*$ and inserts a tuple $\langle J_i,h_{i}\rangle$ to $L_v$. Finally, it outputs $h_{i}$ as the answer.
\end{itemize}

\item
\textbf{Extract-Queries:} $\mathcal{A}_1$ makes the query on $\langle id_i, Q_{1i} \rangle$.
\begin{itemize}
\item[-]
Firstly, $\mathcal{C}$ recovers the corresponding $\langle I_i, c_i,\alpha_i, Q_{2i} \rangle$ from the list $L_c$. If $c_i = 0$, $\mathcal{C}$ returns failure. If $c_i = 1$ and $L$ contains $\langle id_i, s_{1i}, S_{2i}, Q_{1i} \rangle$, $\mathcal{C}$ checks if $S_{2i}=\perp$.
\item[-]
If $S_{2i}\neq\perp$, $\mathcal{C}$ returns the current $S_{2i}$ to $\mathcal{A}_1$. Otherwise, $H_1(id_i, Q_{1i})$ is set as $\alpha_iP$. $\mathcal{C}$ computes $S_{2i}=\alpha_iQ_{KGC}$, then $\mathcal{C}$ inserts a tuple $\langle id_i, s_{1i}, S_{2i}, Q_{1i} \rangle$ to the list $L$ and outputs $S_{2i}$ as the answer.
\item[-]
Again, if $c_i = 1$, the list $L$ does not contain $\langle id_i, s_{1i}, S_{2i}, Q_{1i} \rangle$. Then, $\mathcal{C}$ sets $S_{2i} = \perp$ and computes $S_{2i}=\alpha_iQ_{KGC}$. Finally, $\mathcal{C}$ inserts a tuple $\langle id_i, s_{1i}, S_{2i}, Q_{1i} \rangle$ to $L$ and replies $S_{2i}$ as output.
\end{itemize}

\item
\textbf{Public-Key-Queries:} $\mathcal{A}_1$ makes the query on an identity $id_i$.
\begin{itemize}
\item[-]
If $\langle id_i, s_{1i}, S_{2i}, Q_{1i} \rangle$ is in $L$, $\mathcal{C}$ checks if $Q_{1i} = \perp$. If holds, $\mathcal{C}$ selects $s_{1i}\in Z_q^*$ and $Q_{1i}= s_{1i}P$. It updates $\langle Q_{1i}, s_{1i}\rangle$ to $L$ and replies $Q_{1i}$ to $\mathcal{A}_1$. Otherwise, $\mathcal{C}$ returns $Q_{1i}$ to $\mathcal{A}_1$.
\item[-]
If $\langle id_i, s_{1i}, S_{2i}, Q_{1i} \rangle$ is not in $L$, let $S_{2i} = \perp$, then selects a random $s_{1i}\in Z_q^*$ and sets $Q_{1i}= s_{1i}P$. $\mathcal{C}$ inserts a tuple $\langle id_i, s_{1i}, S_{2i}, Q_{1i} \rangle$ to $L$ and replies $Q_{1i}$ to $\mathcal{A}_1$.
\end{itemize}

\item
\textbf{Secret-Key-Queries:} $\mathcal{A}_1$ makes the query on an identity $id_i$.
\begin{itemize}
\item[-]
If $\langle id_i, s_{1i}, S_{2i}, Q_{1i} \rangle$ is in $L$, $\mathcal{C}$ checks if $s_{1i} = \perp$. If holds, $\mathcal{C}$ selects a random $s_{1i}\in Z_q^*$. It also returns $s_{1i}$ and adds tuple $\langle id_i, s_{1i}, S_{2i}, Q_{1i} \rangle$ to the list $L$. Otherwise, $s_{1i}\neq\perp$, $\mathcal{C}$ replies $s_{1i}$ to $\mathcal{A}_1$.
\item[-]
If $\langle id_i, s_{1i}, S_{2i}, Q_{1i} \rangle$ is not in $L$, $\mathcal{C}$ sets $s_{1i} = \perp$ and replies a random $s_{1i}\in Z_q^*$ to $\mathcal{A}_1$.
\end{itemize}

\item
\textbf{Replace-Public-Key queries:} $\mathcal{A}_1$ chooses new public key $Q_{1i}'$ for an identity $id_i$.
\begin{itemize}
\item[-]
If $\langle id_i, s_{1i}, S_{2i}, Q_{1i} \rangle$ is in $L$, $\mathcal{C}$ sets $Q_{1i} = Q_{1i}'$ and $s_{1i}=\perp$. It updates a tuple $\langle id_i, s_{1i}, S_{2i}, Q_{1i} \rangle$ to the list $L$.
\item[-]
If $\langle id_i, s_{1i}, S_{2i}, Q_{1i} \rangle$ is not in $L$, $\mathcal{C}$ sets $Q_{1i} = Q_{1i}'$ and $s_{1i}=\perp$, then it inserts a tuple $\langle id_i, s_{1i}, S_{2i}, Q_{1i} \rangle$ to the list $L$.
\end{itemize}

\item
\textbf{CLAS-Sign-Queries:} In this queries, $\mathcal{C}$ provides valid signatures of any message $m_i$ of $id_i$ with list $L = \langle id_i, s_{1i}, S_{2i}, Q_{1i} \rangle$, $L_c = \langle I_i, c_i,\alpha_i, Q_{2i} \rangle$, $L_v = \langle J_i, h_{i}\rangle$, and answers the query as follows:
\begin{itemize}
\item[-]
If $L$ is not empty and $c_i=1$, $\mathcal{C}$ checks if $s_{1i}=\perp$. If $s_{1i}=\perp$, $\mathcal{C}$ makes \textbf{Public-Key-Queries} to generate $s_{1i}$ and $Q_{1i}=s_{1i}P$.
\item[-]
If $L$ is empty, $\mathcal{C}$ makes \textbf{Public-Key-Queries} to generate $s_{1i}$ and $Q_{1i}=s_{1i}P$ and adds them to list $L$.
\item[-]
$\mathcal{C}$ tries to generate the signature. If $c_i = 1$, $\mathcal{C}$ returns failure. Otherwise, $\mathcal{C}$ picks a random $k_i\in Z^*_q$, and computes
\begin{equation} \label{eqn4}
  \begin{split}
  V_i&= k_iQ_{1i} \\
  h_i&= H_2(m_i,V_i) \\
  U_i&= S_{2i}+k_ih_is_{1i}Q_{KGC}
  \end{split}
  \end{equation}

\item[-]
Output $\sigma_i=\langle U_i,V_i \rangle$ as the signature on $m_i$.
\end{itemize}
It is easy to verify $\sigma_i$ via the above equation, so the simulation is perfect. If $\mathcal{C}$ does not abort this game, none can distinguish the simulation from a legal signer.
\end{itemize}

Eventually, with nonnegligible probability, $\mathcal{C}$ obtains two valid signatures  $\sigma_i=\langle U_i, V_i \rangle$ and  $\sigma_i'=\langle U_i', V_i \rangle$ with help of $\mathcal{A}_1$, where $U_i\neq U_i'$. Then, we $\mathcal{C}$ the following two equations:
\begin{equation}
U_i=S_{2i}+k_ih_is_{1i}Q_{KGC} \label{eq1}
\end{equation}
\begin{equation}
U_i'=S_{2i}+k_ih_i's_{1i}Q_{KGC}\label{eq2}
\end{equation}

Multiplying both side of equation \eqref{eq1} with $(h_i)^{-1}$ and both side of equation \eqref{eq2} with $(h_i')^{-1}$, we can obtain \eqref{eq3} and \eqref{eq4}

\begin{equation}
(h_i)^{-1}U_i=(h_i)^{-1}S_{2i}+(h_i)^{-1}h_ik_is_{1i}Q_{KGC} \label{eq3}
\end{equation}

\begin{equation}
(h_i')^{-1}U_i'=(h_i')^{-1}S_{2i}+(h_i')^{-1}h_i'k_is_{1i}Q_{KGC}\label{eq4}
\end{equation}

Subtract \eqref{eq4} from \eqref{eq3}
\begin{equation}
(h_i')^{-1}U_i'-(h_i)^{-1}U_i=[(h_i')^{-1}-(h_i)^{-1}]S_{2i}\label{eq5}
\end{equation}

Then, $\mathcal{C}$ obtains $\langle id_i, s_{1i}, S_{2i}, Q_{1i} \rangle $ in $L$ and $\langle I_i, c_i,\alpha_i, Q_{2i} \rangle$ in $L_c$, respectively. If $c_i = 1$, $\mathcal{C}$ aborts. Otherwise, if $c_i = 0$, $Q_{2i}=\alpha_ibP$, now $Q_{KGC} = aP = s_{KGC}P$. Because of $S_{2i}=aQ_{2i}=a\alpha_ibP$, we can obtain \eqref{eq6} and \eqref{eq7} as follows:
\begin{equation}
(h_i')^{-1}U_i'-(h_i)^{-1}U_i=[(h_i')^{-1}-(h_i)^{-1}]ab\alpha_iP\label{eq6}\\
\end{equation}
\begin{equation}
abP = [(h_i')^{-1}U_i'-(h_i)^{-1}U_i][((h_i')^{-1}-(h_i)^{-1})\alpha_i]^{-1}\label{eq7}
\end{equation}

Therefore, $\mathcal{C}$ finds $abP$ as the solution to CDHP and solves CDHP with the probability

$$\varepsilon' \geq \frac{1}{(q_k+1)e}\varepsilon$$

There are three events needed by $\mathcal{C}$ to succeed:
$E_1$ is the result of any Extract-Queries  raised by $\mathcal{A}_1$ does not abort.
$E_2$ represents $\mathcal{A}_1$ generates a valid signature that can be verified.
$E_3$ represents the probability that $\mathcal{A}_1$ outputs a valid forgery and $\mathcal{C}$ does not leave the game.
The probability of $\mathcal{C}$ success is that all the three events mentioned above happen:
$$P[E_1\wedge E_2\wedge E_3] = P[E_1]P[E_2|E_1]P[E_3|E_2\wedge E_1]$$.

\begin{itemize}
\item Claim 1: The probability of $E_1$ happening is at least $(1-\lambda)^{q_k}$, because $P[c_i=1]=(1-\lambda)$ and it takes at leat $q_k$ queries. So, $P[E_1] \geq (1-\lambda)^{q_s}$.
\item Claim 2: The Probability that $E_2|E_1$ happens is at least $\varepsilon$. So $P[E_2|E_1]\geq \varepsilon$
\item Claim 3: The probability that $E_3|E_2\wedge E_1$ happens is at least $\lambda$, because $P[c_i=0]=\lambda$, and $E_1|E_2$ both happen. So $P[E_3|E_2\wedge E_1]\geq \lambda$
\end{itemize}

Therefore, we can conclude that the probability of all three events happening is as follows:
\begin{equation*}
    \begin{array}{rcl}
       P[E_1\wedge E_2\wedge E_3] &=& P[E_1]P[E_2|E_1]P[E_3|E_2]\\
                   & = & \lambda(1-\lambda)^{q_k}\varepsilon \\
    \end{array}
\end{equation*}

We suppose $\lambda = \frac{1}{(q_k+1)}$. Then,

$$\varepsilon'  \geq \lambda(1-\lambda)^{q_k}\varepsilon$$
$$\varepsilon'  \geq \frac{1}{(q_k+1)}[1- frac{1}{(q_k+1)}]^{q_k}\varepsilon$$

If $q_k$ is sufficiently large, $[1-\frac{1}{q_k+1}]^{q_k}$ tends to $\frac{1}{e}$. So, the final probability is as follows:

$$\varepsilon'  \geq \frac{1}{(q_k+1)e}\varepsilon$$

A forged aggregate signature could be generated in the following way by $\mathcal{C}$:

\begin{equation} \label{eqn12}
  \begin{split}
  U &= \sum_{i=1}^n U_i \\
  V &= \sum_{i=1}^n h_iV_i
  \end{split}
  \end{equation}
\textbf{Theorem 2.} The proposed CL-AS scheme is existentially unforgeable against the second kind of adversary $\mathcal{A}_2$ assuming the CDHP is hard.

\textbf{Proof.} This security property also relies on the hardness of CDHP. Assuming the CDHP is intractable, we can prove that our scheme is secure in the similar way in Theorem 1. Thus, we omit the proof in detail.

\section{A Large-scale Concurrent Data Anonymous Batch Verification Scheme for MHCS}

Due to the unique security requirements of mobile healthcare crowd sensing, we design an anonymous batch verification scheme for large-scale concurrent data. It can provide privacy-preserving batch verification of the uploaded health data in MHCS and achieve multi-user access authentication.

\subsection{Scheme Description}

The proposed scheme consists of five algorithms, such as: \emph{Initialization}, \emph{Registration}, \emph{Signing}, \emph{Anonymous Aggregation}, and \emph{Batch Verification}. Here, we list some notations in Table \ref{tabl:notation} to facilitate our understanding. Then, we give the assumption of the time synchronization between the requested DC and MHCS participants. The proposed scheme is introduced as follows:

\begin{itemize}
\item[1)] \emph{Initialization.} MS establishes an enrollment system as follows:
\begin{itemize}
\item[a.] MS define $G_a$ as a additive group, $G_m$ as a multiplicative group, $q$ as the order, $P$ as the generator of $G_a$, $e: G_a \times G_a \rightarrow G_m $ as a bilinear map, $H_1 : {\{0, 1\}}^\ast\times G_a \rightarrow G_a$ and $H_2 : {\{0, 1\}}^\ast\times G_a \rightarrow Z^*_q$ as two secure hash functions.
\item[b.] Given $l$, MS selects its private key $s_{MS}$ randomly and calculates its public key $Q_{MS} = s_{MS}P$. Then, it opens the system parameters $\langle l, q, P, G_a, G_m, e, H_1, H_2, Q_{MS} \rangle$. We suppose that DC regards $\langle s_{DC}, Q_{DC} \rangle$ as its long-term key pair, where $Q_{DC} = s_{DC}P$.
\end{itemize}

\item[2)] \emph{Registration.} A participant and the MS perform the following steps to access a DC as follows:
\begin{itemize}
\item[a.] The participant, marked as $C_i$, chooses a random number $s_{1i} \in Z_q^*$ as the half private key, and it obtains $S_{2i}$ from MS who computes $S_{2i}= s_{MS}Q_{2i}$, where $Q_{2i} = H_1(id_i,Q_{1i})$ as the other half part private key. $C_i$ sets $\langle s_{1i}, S_{2i}\rangle$ as its private key. Then, it sends $\langle id_i,Q_{1i} \rangle$ to MS.
\item[b.] Upon receiving $\langle id_i,Q_{1i} \rangle$, MS chooses a random number $a_i \in Z_q^*$ and calculates
\begin{equation} \label{eqn13}
  \begin{split}
  Q_{2i}     & = H_1(id_i,Q_{1i}) \\
  S_{2i}     & = s_{MS}Q_{2i} \\
  index_{si} & = a_iS_{2i} \\
  index_{vi} & = a_iQ_{2i}
  \end{split}
  \end{equation}
Thus, MS stores serial number $sn_i = \langle id_i, Q_{1i}, Q_{2i},$
$index_{si}, index_{vi} \rangle$. Then, it sends $SN_i = index_{vi}$ and $index_{si}$ to the participant with $id_i$. All of the registration information should be transmitted via a secure channel.
\end{itemize}

\begin{figure}[ht]
  \centering
  \includegraphics[width=8.5cm]{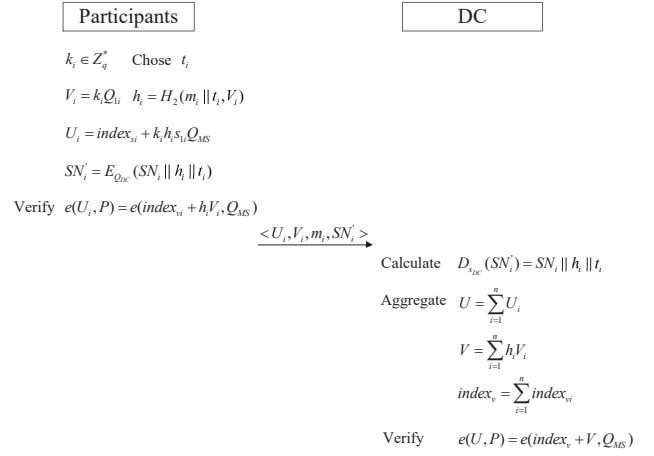}\\
  \caption{The flowchart of the concurrent data anonymous batch verification scheme}\label{Flowchart}
\end{figure}

\item[3)] \emph{Signing.} $C_i$ chooses a random number $k_i\in Z^*_q$ and a time stamp $t_i$, where $t_i$ is the system time to maintain the freshness of the message, and calculates
\begin{equation} \label{eqn14}
  \begin{split}
  V_i   &= k_iQ_{1i} \\
  h_i   &= H_2(m_i||t_i,V_i) \\
  U_i   &= index_{si} + k_ih_is_{1i}Q_{MS} \\
  SN_i' &= E_{Q_{DC}}(SN_i||h_i||t_i)
  \end{split}
  \end{equation}
  Each required sensing data could be verified by
 \begin{equation}\label{eqn15}
  e(U_i,P)=e(index_{vi}+h_iV_i, Q_{MS})
 \end{equation}
  respectively. Then, $C_i$ uploads $\langle U_i, V_i, m_i, SN_i' \rangle$ to DC who issues the sensing task. Additionally, we can easily prove the correctness of the equation \eqref{eqn15} as follows:
  \begin{eqnarray*}
  e(U_i,P)&= &e(index_{si}+k_ih_is_{1i}Q_{MS},P)\\
          &= &e(index_{si},P)e(k_ih_is_{1i}Q_{MS},P)\\
          &= &e(a_{i}S_{2i},P)e(k_ih_is_{1i}P,Q_{MS})\\
          &= &e(a_{i}s_{MS}Q_{2i},P)e(k_ih_iQ_{1i},Q_{MS})\\
          &= &e(index_{vi},Q_{MS})e(h_iV_i,Q_{MS})\\
          &= &e(index_{vi}+h_iV_i,Q_{MS})
  \end{eqnarray*}
\item[4)] \emph{Anonymous Aggregation.} DC plays a role of the aggregator to merge all collected authentication information of different participants to a single verification message. Upon receiving $\langle U_i, V_i, m_i,$ $SN_i' \rangle$, DC calculates $D_{s_{DC}}(SN_i') = SN_i||h_i||t_i$. For an aggregate set of $n$ participants ${C_1, C_2, ......, C_n}$ and a set of signatures $\langle U_i, h_iV_i \rangle$, when the time T is up, DC aggregates all the received signatures as follows:
\begin{equation} \label{eqn16}
  \begin{split}
  U       &= \sum_{i=1}^n U_i \\
  V       &= \sum_{i=1}^n h_iV_i \\
  index_v &= \sum_{i=1}^n index_{vi}
  \end{split}
  \end{equation}
Then, DC treats $\sigma = \langle U,V,index_v \rangle$ on all health data $\langle m_1, m_2, \ldots,$ $m_n \rangle$ as the aggregated authentication message.

\item[5)] \emph{Batch Verification.} As illustrated in Fig. \ref{Flowchart}, DC verifies the validity of $e(U,P)$ $= e(index_v+V,Q_{MS})$. If the equation holds, DC approves all health data uploaded by participants within the time slot T as legal data. Otherwise, DC aborts this procedure. Here, DC can verify the validity of the equation as follows:
    \begin{eqnarray*}
        e(U,P) & = & e(\sum\limits_{i=1}^{n} U_i,P)\\
               & = & \prod\limits_{i=1}^{n}e(U_i,P)\\
               & = & \prod\limits_{i=1}^{n}e(index_{si}+k_ih_is_{1i}Q_{MS},P)\\
               & = & \prod\limits_{i=1}^{n}e(index_{si},P)e(k_ih_is_{1i}Q_{MS},P)\\
               & = & e(index_v,Q_{MS})e(V,Q_{MS})\\
               & = & e(index_v+V,Q_{MS})
    \end{eqnarray*}
\end{itemize}

\subsection{Security Analysis}
For convincing, we analyze the security of the large-scale concurrent data anonymous batch verification scheme in this part.

\begin{table*}[ht]
\scriptsize
 \caption{COMPLEXITY COMPARISON BETWEEN DIFFERENT SCHEMES }\label{perf-com}
\begin{center}
\setlength{\extrarowheight}{0.3cm}
\begin{tabular}{c|c|c|c|c}
  \hline
 \raisebox{0.1cm}[0pt]{Scheme} & \quad \raisebox{0.1cm}[0pt]{Signing} & \quad \raisebox{0.1cm}[0pt]{Verification} & \quad \raisebox{0.1cm}[0pt]{Aggregation} & \quad \raisebox{0.1cm}[0pt]{Aggregate Verification}  \\

  \hline
  \raisebox{0.1cm}[0pt]{THH} & \quad \raisebox{0.1cm}[0pt]{4nH+3nS} & \quad \raisebox{0.1cm}[0pt]{5nH+4nP+2nS} & \quad \raisebox{0.1cm}[0pt]{0}& \quad \raisebox{0.1cm}[0pt]{4P+2nS}  \\
  \hline
  \raisebox{0.1cm}[0pt]{Malhi-Batra} & \quad \raisebox{0.1cm}[0pt]{nH+4nS} & \quad \raisebox{0.1cm}[0pt]{2nH+3nP+3nS} & \quad \raisebox{0.1cm}[0pt]{0} & \quad \raisebox{0.1cm}[0pt]{3P+3nS} \\
  \hline
  \raisebox{0.1cm}[0pt]{XGCL} & \quad \raisebox{0.1cm}[0pt]{nH+3nS} & \quad \raisebox{0.1cm}[0pt]{2nH+3nP+2nS} & \quad \raisebox{0.1cm}[0pt]{ 0 }   & \quad \raisebox{0.1cm}[0pt]{3P+2nS} \\
  \hline
\raisebox{0.1cm}[0pt]{Ours}  & \quad \raisebox{0.1cm}[0pt]{nH+2nS}  & \quad \raisebox{0.1cm}[0pt]{2nH+2nP+nS} & \quad \raisebox{0.1cm}[0pt]{2nS}  & \quad \raisebox{0.1cm}[0pt]{2P}   \\
  \hline

\end{tabular}
\end{center}
\end{table*}

\textbf{Theorem 3.} The proposed scheme satisfies batch authentication, non-repudiation, and anonymity.

\textbf{Proof.} We will give the proof as follows:
\subsubsection{Batch authentication} The proposed scheme is secure due to the intractability of the CDHP. So DC can authenticate the identities of MHCS participants by their signatures on health data. Meanwhile, it can aggregate all signatures from large-scale participants to a single verification message and verify the message by checking $e(U,P) = e(index_v+V,Q_{MS})$. Thus, our scheme can achieve anonymous batch verification.
\subsubsection{Non-repudiation} In our scheme, MHCS participant cannot deny that he/she has submitted the health data. DC can verify his/her signature via the corresponding public key. Then, MS can find serial number $sn_i$ according to the public key and obtain the real identity of the participant.
\subsubsection{Anonymity} In the phase of aggregate verification, due to the distribution of $SN_i$ is random, DC cannot get the real identity of the MHCS participant from $SN_i$. Therefore, even if the opponent has unlimited computing power, it is unable to guess the actual participant's identity with the nonnegligible advantage. Thus, the proposed scheme achieves anonymity.

\section{Performance Evaluate}
In this section, we evaluate the performance of the proposed scheme in two aspects, including computation overhead and storage overhead. Firstly, comparing our scheme with other three existing schemes, we assess the performance of the computation overhead in terms of the computation complexity and time overhead on signing, anonymous aggregation and batch verification. Then, we analyze the storage overhead of the proposed scheme.

\subsection{Computation Overhead}
\subsubsection{Computation Complexity}
We select three existing schemes \cite{tu2014reattack, malhi2015efficient, xiong2013efficient} to compare the computation complexity with our scheme. Due to the computation overhead is mostly caused by three basic cryptographic operations, so we mainly focus on the time consumption of these operations. Here, we only count on computation consumption, while the pre-computation efforts are omitted. We define $P$ as a pairing operation, $S$ as a scalar multiplication in $G_a$ and $H$ as hash functions.

Table \ref{perf-com} shows the complexity comparison between different schemes. We find that, in the signing stage, our scheme only requires $nH+2nS$ operations, while the schemes in \cite{tu2014reattack, malhi2015efficient, xiong2013efficient} require $4nH +3nS$, $nH +4nS$ and $nH +3nS$ respectively. In the verification stage, our scheme needs $2nH+2nP+nS$ operations, rather than $5nH+4nP+2nS$ in \cite{tu2014reattack}, $2nH+3nP+3nS$ in \cite{malhi2015efficient} and $2nH+3nP+2nS$ in \cite{xiong2013efficient}. In addition, in aggregation stage, only our scheme needs $2nS$ scalar multiplications, but it only requires two pairing operations in aggregate verification stage. Hence, compared with the schemes in \cite{tu2014reattack, malhi2015efficient, xiong2013efficient}, our scheme has the least total computation overhead in all four stages -- signing, verification, aggregation and aggregate verification.

\begin{figure*}[ht]
\centerline{
\subfigure[``H'' operation vs. the number of participants]{\includegraphics[width=2.3in]{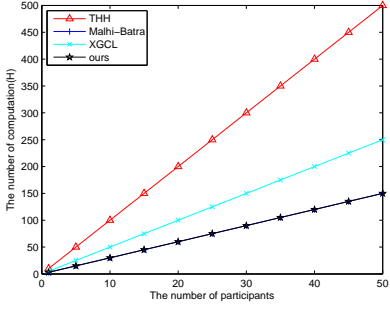}%
\label{Computation_Comparison_H}} %
\subfigure[``P'' operation vs. the number of participants]{\includegraphics[width=2.3in]{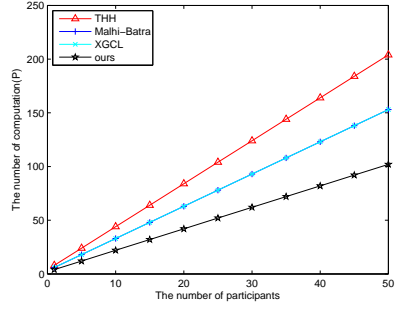}%
\label{Computation_Comparison_P}} %
\subfigure[``S'' operation vs. the number of participants]{\includegraphics[width=2.3in]{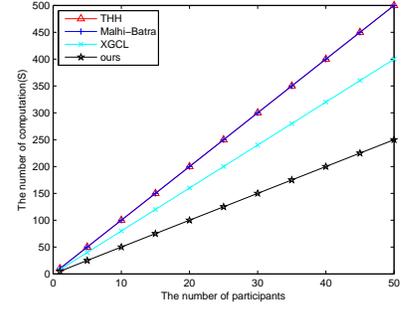}%
\label{Computation_Comparison_S}} %
}
\setlength{\belowcaptionskip}{-2 mm}
\caption{Comparison of computation cost between different schemes}
\label{Computation_Comparison}
\vspace{-1mm}
\end{figure*}

Meanwhile, Fig. \ref{Computation_Comparison} shows the comparison of computation cost between different schemes. And we also find that our scheme has lowest computation complexity than the other schemes \cite{tu2014reattack, malhi2015efficient, xiong2013efficient}, with the increasing of the number of participants. As a whole, our scheme achieve the best performance of the computation complexity.

\subsubsection{Time Overhead}
In order to evaluate and test the performance of time overhead on our scheme, we compare our scheme with other three schemes \cite{tu2014reattack, malhi2015efficient, xiong2013efficient}. For quantitative analysis, we first construct a simulation platform to measure the time overhead. The simulation environment is Ubuntu OS over an Inter Pentium 2.1 GHz processor. We choose type A curve in the Pairing-Based Cryptography (PBC) library -- $y^{2}=x^{3}+x$, to complete the simulation. Here, we assume that $n$ participants try to upload their health data in a certain time slot T.

\begin{figure*}[ht]
\centerline{
\subfigure[Time cost on signing]{\includegraphics[width=2.3in]{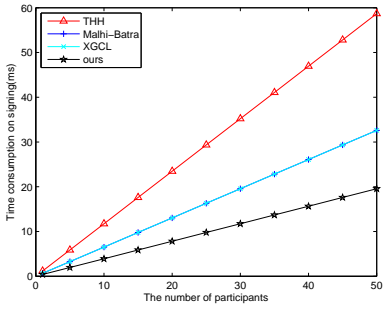}%
\label{Time_Comparison_Sign}} %
\subfigure[Time cost on verification]{\includegraphics[width=2.3in]{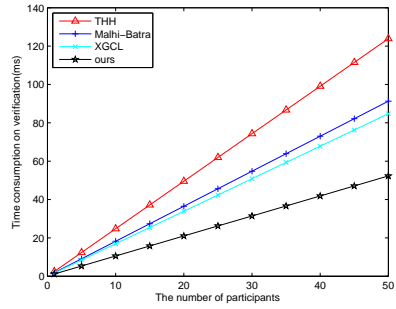}%
\label{Time_Comparison_Verify}} %
\subfigure[Time cost on Aggregation]{\includegraphics[width=2.3in]{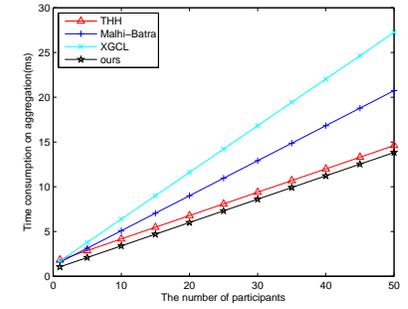}%
\label{Time_Comparison_Aggregation}} %
}
\setlength{\belowcaptionskip}{-2 mm}
\caption{Comparison of time consumption between different schemes}
\label{Simulation_Time}
\vspace{-1mm}
\end{figure*}

Next, we view aggregation as the integration of aggregation and aggregate verification. Then, we record the start time from the beginning of the signing stage to simulate these schemes. Therefore, we can obtain the time overhead of different schemes as shown in Fig. \ref{Simulation_Time}. Compared with the schemes in \cite{tu2014reattack, malhi2015efficient, xiong2013efficient}, the proposed scheme can save 50\%, 42.1\%, 39\% running time respectively.

\begin{figure}[ht]
  \centering
  \includegraphics[width=3.5in]{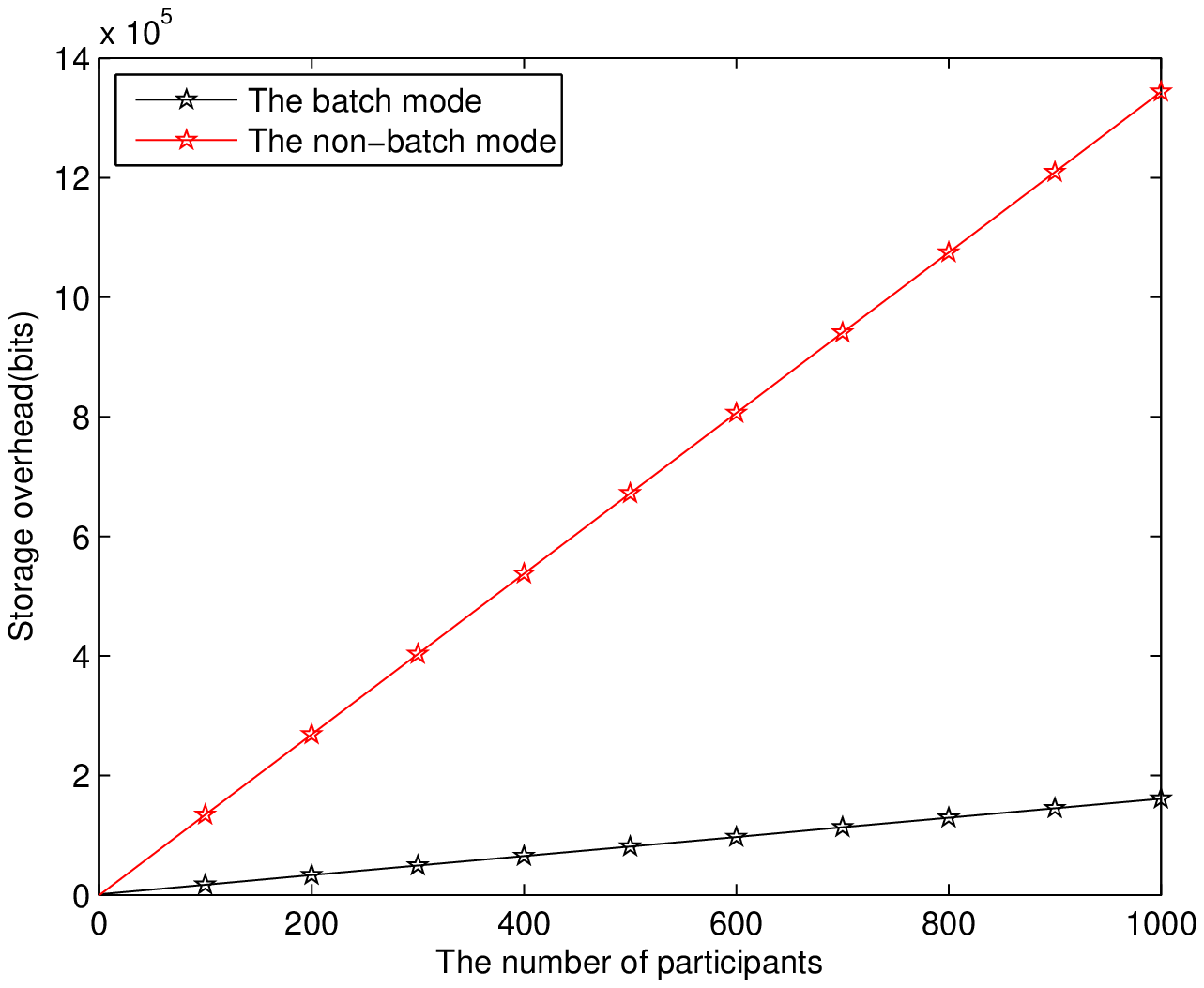}\\
  \caption{The storage overhead of our scheme}\label{storage overhead}
\end{figure}

\subsection{Storage Overhead}
In the proposed scheme, the Data Center (DC) needs to store all collected authentication information of different participants continuously until batch verification is done. Meanwhile, as the aggregator, DC can, in real time, merge the collected authentication information into a single verification message, due to the advantage of the equation (\ref{eqn16}). When time T is up, DC can verify these data in batch. Therefore, the storage overhead of the proposed scheme can be reduced differently according to the number of MHCS participants. For quantitative analysis, we adopt the type A curve with base field size of 512 bits, the cyclic group order of 160 bits, and the embedding degree 2. So, $U_i = 512$ bits, $V_i = 512$ bits, and $SN_i'=160$ bits. Here, we assume that the size of health data $m_i$ is 160 bits as \cite{ren2007broadcast}.

As mentioned before, the verification information of the participant $i$ is $\langle U_i, V_i, m_i,$ $SN_i' \rangle$. Therefore, the corresponding storage overhead of the authentication data is $SO_i = 512+512+160+160 = 1344$ bits. Here, $SO_i$ denotes the storage overhead of the participant $i$. When the time T is up, the total storage overhead of the $n$ participants in this time slot is $SO = 512+512+160n+160 = 160n+1184$ bits. Otherwise, if the verification stage does not utilize the scheme in the batch mode, the total storage overhead of the $n$ participants is $SO' = 1344n$ bits. For better demonstration, we depict the storage overhead on the aforementioned two cases in Fig. \ref{storage overhead}. Then, we can conclude that the storage overhead is greatly reduced in the batch mode.

For all above, the proposed scheme achieves a better performance in terms of computation overhead and storage overhead. It is efficient and suitable for mobile healthcare crowd sensing.

\section{Conclusion}

In this paper, based on an improved CL-AS algorithm, we design an anonymous batch verification scheme for large-scale concurrent data in MHCS scenarios. It meets the EUF-CMA security in the random oracle model based on the intractability of the CDHP. And it can achieve three properties including batch authentication, non-repudiation, and anonymity. Moreover, our scheme can be deployed in MHCS system to offer batch health data authentication and privacy preservation simultaneously. Through quantitative performance analysis, we find that the proposed scheme achieves lower computation overhead and provides better efficiency compared with the existing schemes, and its storage overhead is also reduced greatly. The proposed scheme is an efficient solution for the MHCS systems.

\ifCLASSOPTIONcaptionsoff
  \newpage
\fi

\bibliographystyle{IEEEtran}
\bibliography{mhcs}

\end{document}